\documentclass[11pt]{article}

\usepackage[french]{babel}
\usepackage[utf8]{inputenc}
\usepackage[T1]{fontenc}
\usepackage{amsmath}
\title{Arithmétique des primes pluri-annuelles}
\author{Olivier Garet}
\begin{document}
\maketitle

Cette courte note étudie les propriétés arithmétiques simples des systèmes d'attribution pluriannuelle des primes, telles qu'elles ont été récemment mises en place dans la population des enseignants chercheurs, avec la PEDR d'une part, avec la composante 3 de la RIPEC d'autre part.

Dans un premier temps, on étudie un premier modèle où toute la population candidate, de manière à mettre en lumière l'effet du caractère pluriannuel de l'attribution:  le taux de réussite à une candidature est très différent de la fréquence avec laquelle un individu est bénéficiaire de la prime.

Dans un second temps, on tient compte du fait qu'une partie de la population n'est pas candidate, avec des influences croisées entre le taux de participation à la compétition et le taux de réussite des individus.

\`A la lumière de statistiques disponibles sur le comportement des agents, nous discuterons la possibilité pour la RIPEC d'entraîner une large part de la population. Nous répondons par la négative et proposons finalement une suppression de la composante~3 de la RIPEC au profit d'une réévaluation de la composante~1 du RIPEC (composante statutaire) -- en attendant une réévaluation indiciaire. 

\section{Un premier modèle}

Dans ce premier modèle, on considère une population qui est systématiquement candidate à la prime dès qu'elle en a la possibilité.\\

Notations:\\

\begin{tabular}{|l|l|}
  \hline
  $n$ & nombre d'années d'attribution de la prime\\ \hline
  $q$ & proportion de la population bénéficiaire de la prime\\ \hline
  $a$ & proportion de la population en 1\up{ère} année de la prime\\ \hline
  $p$ & taux de succès dans la sélection d'une année\\ \hline
\end{tabular}  
\quad {}\\

\vspace{0.3cm}

En régime stationnaire, ces quantités sont liées par deux équations simples:\\

D'abord,
\begin{align}
  a&=\frac{q}{n}.
\end{align}

En effet, les bénéficiaires sont également répartis entre les première, deuxième,\dots, $n$-ième année de la prime.\\

On a également l'équation de stationnarité:

\begin{align}
  \label{station}
  a&=(1-(n-1)a)p.
\end{align}

En effet, pour être en première année de bénéfice de la prime, il faut à la fois être susceptible de candidater et être sélectionné.
Les personnes qui ne peuvent pas candidater sont celles qui sont en années $1$, $2$,\dots,$n-1$ de la prime: elles représentent une proportion $(n-1)a$ de la population totales: les candidats sont donc en proportion $1-(n-1)a$, d'où l'équation~\eqref{station}.

C'est une relation du premier degré entre les paramètres $a$ et $p$, que l'on peut résoudre afin d'afficher un paramètre en fonction de l'autre. On a

\[ p=\frac{a}{1-(n-1)a}\text{ et } a=\frac{p}{1+(n-1)p}\]

Naturellement, on préfère plutôt exprimer les choses en fonction du paramètre $q$, qui a une signification politique, ce qui nous donne

\[\boxed{p=\frac{q/n}{1-\frac{n-1}n q}\text{ et } q=\frac{np}{1+(n-1)p}.}\]

Une dernière quantité a une signification économique et politique, c'est la proportion $c$ de la population qui dépose un dossier chaque année.
Cette variable est associée à un coût financier et humain, puisque les dossiers doivent être écrits et doivent être évalués.

On a l'équation

\[ \boxed{c = 1-(n-1)a=1-\frac{(n-1)q}{n}}\]

\section{Application à la PEDR et à la C3 de la RIPEC}

Dans le système de la PEDR, la prime est attribuée pour 4 ans. On a donc $n_{\text{PEDR}}=4$.
Dans une université comme l'université de Lorraine, les Lignes Directrices de Gestion stipulent que l'université attribuera exactement la prime aux personnes qui ont obtenu la note A ou B de la part de l'instance nationale d'évaluation (le CNU). Cette instance attribue les notes suivant des contingents fixés par la loi: $0,2$ pour la note A, $0,3$ pour la note $B$. On a donc $\boxed{p_{\text{PEDR}}=0,5}$, puis

\[ \boxed{q_{\text{PEDR}}=\frac{np}{1+(n-1)p}=\frac{4\times 0,5}{1+3\times 0,5}=0,8}\]

et
\[ \boxed{c_{\text{PEDR}}=1-\frac{(n-1)q}n=0,4}.\]

La valeur trouvée pour $q_{\text{PEDR}}$ est étonnante et ne correspond pas à l'idée qu'on se fait d'une compétition réputée sélective. On y reviendra ultérieurement. En première instance, il faut avoir présent à l'esprit qu'il y avait une grosse auto-censure et que la population qui participe au concours est bien plus restreinte que la population des enseignants-chercheurs.

Parce qu'elle vise dès le départ à être accessible à toute la population des enseignants-chercheurs, la politique de fonctionnement de la C3 du RIPEC doit d'emblée maîtriser le coût financier du versement de la prime. Chaque université détermine sa politique, dans le respect de la loi et des lignes directrices de gestion (LDG) ministérielles. Le protocole LPR donne une valeur de référence qui n'oblige pas l'université, mais est utilisée comme référence dans la dotation des universités.
Cette valeur de référence est $q_{\text{C3}}=0,45$, ce qui amènerait, au taux de succès, si tout le monde était candidat, à
\[p_{\text{C3}}=\frac{q/n}{1-\frac{n-1}n a}=\frac{0,1125}{1-0,75\times 0,45}\approx0,17\]
et
\[c_{\text{C3}}=1-\frac{n-1}{n}q=1-0,75\times 0,45\approx0,66.\]

On voit bien que le résultat est absurde: on aurait un système avec un taux de rejet important, mais qui finalement donnerait la prime à près de la moitié des gens, ce qui conduirait les collègues à passer leur temps à faire et à évaluer des dossiers.

\section{Un modèle plus réaliste}

Chacun aura remarqué la limite évidente du modèle, qui est que tout le monde ne candidate pas aux primes. Le modèle précédent est donc valable dans une sous-population, qui s'estime légitime à participer au concours. Si $N$ est le nombre total d'enseignants chercheurs et $N_{\text{part}}$ l'effectif de cette sous-population, le rapport
\[i=\frac{N_{\text{part}}}{N}\]
  représente donc la proportion des EC de l'établissement intégrés dans la dynamique du concours.

La capacité budgétaire de l'\'Etablissement, sa politique générale, l'amène à déterminer

\[f=\frac{N_{\text{primes}}}{N}\]
qui est la proportion des bénéficiaires de la prime.
Chaque année, le nombre de primes mises au concours est donc
\[N_{\text{primes ann.}}=\frac{N_{\text{primes}}}{n}=\frac{Nf}{n}.\]  

Le taux d'accès de la population sous-candidate à la prime est
\[q=\frac{N_{\text{primes}}}{N_{\text{cand}}}=\frac{N_{\text{primes}}/N}{N_{\text{cand}}/N}=\frac{f}{i}.\]

Pour une valeur de $n$ fixé, les variables $p,q,f$ (qui sont évidemment liées) déterminent la politique de l'employeur, qui peut choisir de baser sa communication publique sur la variable $f$ ou sur la variable $p$. Ces variables sont liées à la variable $i$ qui représente l'inclusion des agents dans le processus de compétition.

Il faut noter que les équations données sont dans le modèle à l'équilibre; les interactions entre les variables se font dans les deux sens. S'il est évident que le taux de succès des candidatures ou le nombre de primes offertes influe sur le nombre de personnes qui acceptent d'entrer dans le processus de candidature, l'inverse peut être vrai: la confiance dans la soutenabilité financière de la politique PEDR pour les universités qui donnaient systématiquement une prime pour les notes $A$ et $B$ est  basée sur l'hypothèse du maintien de l'autocensure des candidatures, héritée de l'ancienne PEDR qui existait avant la LRU.
En effet, comme on l'a vu, si tout le monde avait candidaté, il aurait fallu payer une PEDR à $80\%$ des enseignants-chercheurs, alors que la PEDR bénéficiait à seulement à peu près $20\%$ des enseignants-chercheurs.
Ceci peut expliquer qu'un certain nombre d'universités, parfois très réputées, aient choisi de sortir du processus national d'évaluation, pour éviter la situation désagréable qu'on connu de nombreux personnels qui n'ont pas eu la PEDR malgré une bonne évaluation nationale (le taux de réussite observé à la PEDR est $44\%$~\cite{pedr20,pedr21}, et pas $50\%$, vraisemblablement pour des questions de budget).

La communication du ministère et des universités insiste sur le fait qu'une prime que touchera $45\%$ des gens remplace une prime que ne touchent que $20\%$. En réalité, la limitation théorique de l'ancien dispositif étant à $80\%$, on peut imaginer qu'on aurait pu faire monter ce taux bien plus haut que $45\%$ en élargissant (comme on le fait avec la RIPEC) la liste des motifs de prime et en donnant le budget nécessaire aux établissements, comme on le fait avec la RIPEC.

On peut donc, sans trop de risque, conjecturer que le but premier de la réforme, n'est pas d'augmenter le nombre de bénéficiaires, mais d'augmenter la valeur de $i$ qui représente la part de la population qui \og joue le jeu\fg.

\section{Inclusion des EC dans le mécanisme de la PEDR} 

On peut essayer d'estimer cette part pour la PEDR.
Suivant la note~\cite{pedr21} de la DGRH de février 2022, 11790 enseignants-chercheurs bénéficiaient alors de la PEDR, alors que, selon le panorama 2021 de la DRH~\cite{panorama}, la filière universitaire et la filière hospitalière cumulent 54293 emplois, ce qui permet d'estimer $\boxed{f_{\text{PEDR}}=22\%}$.
Comme $\boxed{p_{\text{PEDR}}=0,44}$, on a $\boxed{q_{\text{PEDR}}=\frac{4\times 0,44}{1+3*0,44}\approx 0,76}$, ce qui mène à l'estimation
\[i_{\text{PEDR}}=\frac{f_{\text{PEDR}}}{q_{\text{PEDR}}}=0,29.\]
Un autre calcul possible pour estimer la population investie dans le dispositif PEDR est de cumuler le nombre de candidats une année donnée avec les 3/4 des bénéficiaires.
Cela donne l'estimation légèrement différente \[i_{\text{PEDR}}=\frac{0.75\times 11790+6740}{54293}=0,34.\]

Les deux nombres sont suffisamment proches pour qu'on retienne qu'un universitaire sur trois était mobilisé dans le mécanisme de candidature à la PEDR.
On peut noter que ces deux estimations sont proches du rapport candidats/promouvables observé au moment des campagnes d'avancement de grade ($34\%$ pour la campagne 2022~\cite{avancement22}).  

\section{Prospective}
La création de la composante 3 du RIPEC entend changer la donne, et le résultat de la première campagne d'attribution des primes va dans ce sens puisque le nombre de candidats à la RIPEC en 2022 était 11500~\cite{ripec22}, soit une augmentation de $70\%$ par rapport au nombre de candidats à la PEDR en 2021 (6740).

Le taux de référence de $0,45$ n'a sans doute pas été choisi au hasard: proche de $0,5$, il laisse entendre qu'on a une chance sur deux de sortir victorieux de l'épreuve, alors que nous avons montré qu'à cause du caractère pluriannuel d'affectation de la prime, avec une cible $f=0,45$, le taux de réussite sur une année si tout le monde est candidat est de $0,17$

Je fais l'hypothèse que le comportement des personnes est davantage influencé par la valeur de $p$ que par la valeur de $f$, car la plupart des gens sont plus affectés par leurs échecs qu'ils ne sont motivés par les perspectives de réussite sur le long terme.
Les valeurs  des taux de promotion, de prime sont peu connues des collègues; c'est l'expérience sensible qui compte et il n'est pas rare de voir des collègues céder au découragement après deux échecs consécutifs aux campagnes de promotion.

Je fais donc l'hypothèse qu'il existe une valeur critique $p_{\text{crit}}$ minimale d'acceptabilité du succès, et que l'inclusion maximale des personnels dans la dynamique de la prime est donnée par l'équation

\[ i_{\text{max,C3}}=\frac{f}{q_{\text{crit}}}=f_{\text{C3}}\left( \frac1{4p_{\text{crit}}}+\frac34   \right).\]

Rappelons que le taux de succès à la PEDR était $0,44$. Si on se reporte aux statistiques 2022 pour les avancements de grade, on a un taux de succès, tous grades et tous corps EC confondus de $0,49$, avec une différence entre les maîtres de conférence ($0,58$) et les professeurs ($0,39$). Il n'est pas étonnant que l'acceptation du risque soit plus grande chez celles et ceux qui occupent les positions les plus élevées.

Si on accepte en première approximation la valeur $p_{\text{crit}}=0,5$, cela donnerait

\[ i_{\text{max,C3}}=1,25 f_{\text{C3}},\text{ soit } i_{\text{max,C3}}=56,25\%\text{ si }f_{\text{C3}}=0,45.\]

Il semble donc clair qu'une part importante des enseignants-chercheurs (plus de $40\%$) se tiendra à l'écart de la C3, que ce soit par choix idéologique ou par dépit.

Après la première année, il semble toutefois très loin d'être acquis que la C3 emmènera avec elle autant de gens:
si on additionne aux 11500 candidats à la C3 en 2022 les 11790 personnes qui étaient titulaires de la PEDR, on arrive à un total de 23290, soit 43\% de la population, ce qui encore inférieur à la cible des 45\% de bénéficiaires espérés.
Si on montait à 55\% de la population investie dans le dispositif concurrenciel (et on en est loin), on afficherait un taux de réussite instantané de près de 82\%, ce qui signifierait l'inanité du processus d'évaluation.

\section{Conclusion}

L'histoire de la C3 n'est pas encore écrite; il y a plusieurs issues possibles, mais aucune n'est très engageante: si les gens participent \og beaucoup \fg, la déception devant l'échec sera vite au rendez-vous, provoquant la mise en retrait de nombreux collègues, et beaucoup de ressentiment. Sinon, les primes se partageront avec un taux de succès important dans une population dont le principal mérite sera d'accepter le principe de mise en compétition. Une troisième possibilité, qui signifierait l'échec du protocole, est que le gouvernement repousse au-delà de la date de 2027 l'objectif des 45\%. 

En réalité, il y a une contradiction inhérente au dispositif de la C3: dans une population dont la culture et les valeurs ne sont pas celles de la compétition permanente, on ne peut pas avoir un système qui soit simultanément discriminant et inclusif. Ainsi, si l'objectif n'est pas modifié, il n'est pas exclu que le système s'effondre en quelques années, faute de candidats.

La fusion de la C3 avec la C1 semble donc clairement l'option la plus sage.

\def\refname{References}
\bibliographystyle{plain}
\bibliography{arithmetique_des_primes}

\end{document}